\documentclass[epj]{svjour}

\usepackage[dvips]{graphicx}

\begin{document}
\title{Quantum dynamics of early Universe}
\author{Sergei~P.~Maydanyuk
\thanks{\emph{E-mail:} maidan@kinr.kiev.ua}%
}  
%
%
\institute{Institute for Nuclear Research, National Academy of Sciences of Ukraine,
prosp. Nauki, 47, Kiev, 03680, Ukraine}
\date{Received: date / Revised version: date}



\abstract{
In order to study quantum dynamics of the FRW-universe of closed type, definitions of velocity, Hubble function and duration of the evolved universe are introduced into cosmology.
The proposed definitions are characterized by high stability of calculations and easy for use.
The introduced characteristics are supported by calculations of wave function in the fully quantum (non-semiclassical) approach.
%
We achieve high precision agreement between the classical and quantum calculations after the formation of Universe with classical spacetime (i.e. Big Bang).
Such an agreement confirms efficiency of the proposed definitions, and
classical-quantum correspondence allows to obtain quantum information before Big Bang,
to study dynamics of evolution of universe in the first stage and later times.
%
\PACS{
      {98.80.Qc}{Quantum cosmology} \and
      {98.80.Bp}{Origin and formation of the Universe, Big Bang theory} \and
      {98.80.Jk}{Mathematical and relativistic aspects of cosmology} \and
      {03.65.Xp}{Tunneling, traversal time, quantum Zeno dynamics}}
} 

\maketitle



\section{Introduction
\label{sec.introduction}}

Data of astronomical observations suggest on speeding up character of expansion of the present universe. This is resent observations of supernova of type Ia (SNe Ia) \cite{Perlmutter.1998.Nature,Perlmutter.1999.AJ,Riess.1998.AJ,Schmidt.1998.AJ,Tonry.2003.AJ}, more recent cosmic microwave background radiation (CMBR) data~\cite{Spergel.2003.AJS,Bennett.2003.AJS,Tegmark.2004.PRD}, clusters of galaxies~\cite{Pope.1999.AJ}, etc.
This has been designed as ``dark energy'' effect known long before (which practically could be known starting from papers \cite{Einstein.1917.SPAW,Einstein.1919.SPAW} of Einstein introduced the cosmological $\Lambda$ term, see also \cite{Sahni.2002.CQG}) and related presence of ``dark matter'' (see \cite{Oort.1932.BAIN,Zwicky.1933.HPA}, also \cite{Dolgov.2012.EPAN,Dolgov.1988.book} for details).
In order to explain this phenomenon, different approaches in cosmology have been intensively developed.
These are $\Lambda$-term models~\cite{Guth.1981.PRD,Saha.2001.PRD,Cardenas.2003.PRD,Padmanabhan.2003.PR,Saha.2004.PRD,Saha.2006.ASS},
$\Lambda$CDM~\cite{Weinberg.1989.MPR},
phantom-type dark energy~\cite{Caldwell.2003.PRL,Setare.2007.EPJC},
quintessence~\cite{Ratra.1988.PRD,Caldwell.1998.PRL,Zlatev.1999.PRL,Sahni.2000.IJMPD,Saha.2006.IJTP},
$k$-essence,
quintom~\cite{Feng.1988.PLB},
standard Chaplygin gas~\cite{Kamenshchik.2001.PLB},
generalized Chaplygin gas~\cite{Bento.1988.PRD,Bouhmadi-Lopez.2008.IJMPD},
modified Chaplygin gas~\cite{Lu.2008.MPLA},
oscillating dark-energy~\cite{Steinhart.2002.PRD},
holographic dark energy~\cite{Li.2004.PLB,Wu.2008.PLB,Setare.2007.PLB},
non-commutativity models~\cite{Bertolami.2008.PRD,Abreu.2012.JHEP}, etc.
If to look at such a variety of models via description of formation of universe and ability to test by astronomical observations, then one can divide them into two groups.

To the first group we can include the models, describing evolution of expansion of the universe at present time. However, these models mainly are not quantum, and they do not study the initial stage in details.
One of the characteristics used in these models is the parameter of Hubble: from comparison of its calculated values with observational data some conclusions are made about the restrictions applied on the considered models (for example, see~\cite{Lu.2009.EPJC}) and, sometimes, information about character of the evolution of the universe at present times is extracted.

To the second group one can include the models maximally oriented on study of the formation of the universe and its further evolution in the first stage (i.e., Big Bang).
Models oriented on study of quantum peculiarities form one research line in such a group. Here, start of expanding of the universe is associated with tunneling transition through the barrier, usually based on quantum cosmology~%
\cite{Fomin.1975.DAN,Zeldovich.1981.SAJ,Vilenkin.1982.PLB,Hartle.1983.PRD,Vilenkin.1983.PRD,Linde.1984.LNC,%
Linde.1984.JETP,Zeldovich.1984.LNC,Rubakov.1984.PLB,Vilenkin.1984.PRD,Atkatz.1984.PRD,Vilenkin.1986.PRD,%
Vilenkin.1988.PRD,Vilenkin.1994.PRD,Vilenkin.1997.PRD.Comments,Bouhmadi-Lopez.2002.PRD,Vilenkin.2003.PRD,Kim.2004.JKPS}.
However, in frameworks of such models it is usually very difficult to hold out the calculations to the present times, and to connect them with data of astronomical observations.

Natural and important interest is dynamics of evolution of universe in the first stage: It could be interesting to know how the universe evolves in details in the light of quantum physics. However, this question has not been studied in needed level that can be explained by serious difficulties.
From literature we find that practically all calculations of quantum rates of universe evolution were performed for the case of $k=1$ only~\cite{AcacioDeBarros.2007.PRD,Monerat.2007.PRD}. In such a case, potential used in the Wheeler-De Witt equation has a barrier, and formation of universe is described as tunneling transition through it.
A basic idea, which quantum physics provides us, is attempt to perform analysis on the basis of wave packet which evolves through this
barrier~\cite{AcacioDeBarros.2007.PRD,Monerat.2007.PRD,Oliveira-Neto.2011.IJMPCS,%
Pedram.2007.IJTP,Pedram.2007.PLB,Pedram.2007.CQG,Pedram.2008.PLB.v659,Pedram.2008.PLB.v660,Pedram.2008.GRG,%
Pedram.2008.PRD,Pedram.2008.JCAP,Pedram.2009.PLB,Pedram.2010.IJTP,Pedram.2010.PLB,%
Correa_Silva.2009.PRD,Vakili.2010.CQG,Majumder.2011.PLB}. The difficulty can be explained by huge non-stability of such calculations, which are started after leaving of wave packet from the barrier region, practically.

Another problem is necessity to include univocal correspondence between such dynamical evolution and initial, boundary conditions, which reinforce such divergencies but can lead to new quantum effects (see \cite{Maydanyuk.2011.EPJP}, for details).
So, construction of new special apparatus of quantum mechanics based on other ideas could be crucial in resolving of such a problem.

As another research line in this second group, string and brane models~%
\cite{Brandenberger.1989.NPB,Park.2000.PRD,Gasperini.2003.PR,Battefeld.2006.RMP,Veneziano.v265.PLB,%
Brustein.2006.PRD,Cline.2006.hep-th.0612129,Bouhmadi-Lopez.2007.JCAP,Lee.2007.CQG,Lee.2008.CQG},
and also multidimensional models~\cite{Carugno.1996.PRD} are intensively investigated.
However, they involve other aspects of physics and geometry.
Quantum formalism in such theories is usually realized via lagrangian approaches of quantum fields where authors try to describe interactions between different ingredients.
But, this way does not provide apparatus on such successful level for study of quantum effects considered above.
Open questions are multi-particle interactions (formed not small collective effects in nuclear physics), quantum non-locality,
role of boundary and initial conditions%
\footnote{As clear example, multi-particle interactions of nuclear forces are absolutely not small and, so, they is serious problem. From literature one can find that, up to present day, microscopic models for description of nuclear forces (based on Feynmann diagrams procedures) in processes with participation of many nucleons (fission, alpha-decay, emission of proton or cluster from nucleus, etc.) is essentially less successful (in description of experimental data), than models based on folding approaches where nuclear system is studied as medium.
Many particle interactions are related with collective dynamics, they can be directly connected with quantum non-locality which can be studied by quantum mechanics methods (see \cite{Maydanyuk.2011.JMP}, as demonstration).}.

The expanded number of papers present calculations in frameworks of the semiclassical approximations (for example, see~\cite{Vilenkin.1988.PRD,Vilenkin.1995,Casadio.2005.PRD.D71,Casadio.2005.PRD.D72,Luzzi.2006.PhD}).
However, for accurate study one would like to renounce such approximations.
Sometimes researchers use exactly solvable potentials (for example, see~\cite{Vilenkin.1988.PRD,Yurov.2001.CQG,Yurov.2004.TMP,Yurov.2005.PRD,Yurov.2009.TMP,Yurov.2011.TMP,Garcia.2006.IJTP}).
But the total number of such initial cases (i.e. potentials generating corresponding hierarchies) can be counted on the fingers, while proper grounds for such research should be tools for work with barriers of arbitrarily shape and without semiclassical approximations.
This has caused interest to development of methods of quantum mechanics specially oriented on quantum cosmology (for example, see~\cite{Maydanyuk.2003.PhD-thesis,Rubakov.2002.PRD,Esposito.2012.JPA}).

One can understand the most clearly the quantum properties of the formation of the universe in the Friedmann-Robertson-Walker (FRW) model. For example, in~\cite{Maydanyuk.2011.EPJP} the resonant behavior of the penetrability of the barrier in dependence on the chosen initial and boundary conditions (i.e. coordinate of the start of wave, position of the external boundary outside the barrier, energy of the radiation) was opened.
This result can strongly (up to several thousand of percents) change the results obtained by other authors in the semiclassical approximation.
Another result of this paper (and \cite{Maydanyuk.2008.EPJC,Maydanyuk.2010.IJMPD}) is that space-time in the first stage of evolution of the universe is discrete rather than continuous. But, in later times this property gradually decreases and is lost. At the same time, the semiclassical methods are not sensitive to such peculiarity.


By motivations above, in this paper we investigate further the fully quantum approach proposed in \cite{Maydanyuk.2011.EPJP}.
For analysis, how fastly the universe is expanding in frameworks of the different models, we introduce quantum definitions of the velocity of expansion of the universe. In order to realize connection of fully quantum calculations with current astronomic observations, we introduce the function of Hubble in quantum formulation. This basis allows us to investigate dynamics in quantum cosmology, to become a basis for testing the quantum models via astronomical observations.



\section{Formulation of quantum dynamics of evolution of universe in Friedmann-Robertson-Walker metric
\label{sec.model}}

\subsection{Model
\label{sec.model.1}}

We shall start from consideration of FRW model.
We write action in the form (for example, like in~\cite{Vilenkin.1995}, Eq.~(1), p.~2, in case without matter term):
\begin{equation}
\begin{array}{cc}
  S = \displaystyle\int \sqrt{-g}\: \biggl( \displaystyle\frac{R}{16\pi\,G} - \rho \biggr)\; dx^{4}, &
  R =
  \displaystyle\frac{6\dot{a}^{2} + 6a\ddot{a} + 6k}{a^{2}},
\end{array}
\label{eq.model.1.1}
\end{equation}
where $t$ and $r$, $\theta$, $\phi$ are time and spherical space coordinates, our metric signature is $(-,+,+,+)$ as in~\cite{Weinberg.1975,Trodden.TASI-2003},
$a$ is the scale factor,
an overdot denotes a derivative with respect on time $t$,
$R$ is Ricci scalar,
$k$ is curvature of the spatial sector which equals to $-1$, $0$ or $+1$ (for open, flat and closed universe, correspondingly).
We shall use system of units in which $c = \hbar = 1$, and define the reduced Plank mass by $M_{\rm P} \equiv (8\pi G)^{-1/2} \simeq 10^{18}$~GeV, $G$ is Newtonian constant.
The energy density in presence of radiation and dark sector can be written as
\begin{equation}
\begin{array}{lcl}
  \rho\,(a) & = & \rho_{\rm gCg}\,(a) + \displaystyle\frac{\rho_{\rm rad}}{a^{4}}, \\

  \rho_{\rm gCg}\, (a) & = &
  \biggl( A + \displaystyle\frac{B}{a^{3\,(1+\alpha)}} \biggr)^{1/(1+\alpha)} = \\
  & = &
  \displaystyle\frac{1}{\pi}\,
  \biggl( \bar{A} + \displaystyle\frac{\bar{B}}{a^{3\,(1+\alpha)}} \biggr)^{1/(1+\alpha)},
\end{array}
\label{eq.model.1.2}
\end{equation}
where $\bar{A} = A\, \pi^{1+\alpha}$ and $\bar{B} = B\, \pi^{1+\alpha}$. Here, the first item $\rho_{\rm gCg}\,(a)$ represents the generalized Chaplygin gas, including dark energy term $A$ and dark matter term $B$, the second term $\rho_{\rm rad}$ describes energy density of radiation.



Substituting (\ref{eq.model.1.2}) to (\ref{eq.model.1.1}), assuming $\sqrt{-g} = \mathcal{N}\, a^{3}$ where $\mathcal{N}$ is lapse function (for example, like in \cite{Yurov.2007.0710.0094v1}), and choosing $\mathcal{N}=1$ for further calculations,
we obtain lagrangian:
\begin{equation}
  L\,(a,\dot{a}) =
  \displaystyle\frac{3\,a}{8\pi\,G}\:
  \biggl(-\dot{a}^{2} + k -
    \displaystyle\frac{8\pi\,G}{3}\; a^{2}\, \rho(a) \biggr).
\label{eq.model.1.3}
\end{equation}
This form for Lagrangian coincides with Eqs.~(1)--(4) in Ref.~\cite{Monerat.2007.PRD} (at choosing $M_{\rm P} \equiv (8\pi G)^{-1/2} \to 1$, and if generalized Chaplygin gas and energy of radiation terms are included in energy density),
coincides with Eq.~(3) in Ref.~\cite{Bouhmadi-Lopez.2005.PRD} (up to normalized factor $2\pi^{2}$ and renormalization of the energy density $\rho$, which is chosen in the generalized Chaplygin gas form),
similar with Eq.~(17) in Ref.~\cite{Mansouri.1999.PRD} (at $D=3$ and after connecting pressure $p$ via energy density $\rho$ on the further basis of equation of state, in that paper multidimensional universe is studied with arbitrary dimension $D$).
For case for the universe of the closed type without dark matter and radiation
($\alpha=0$, $k=1$, $\mathcal{N}=1$) we obtain
%
%
Eq.~(11) in~\cite{Vilenkin.1995} (up to normalized factor).
Defining hamiltonian as
\begin{equation}
\begin{array}{ccl}
\vspace{1mm}
  h\,(a,p_{a}) & = & p\,\dot{a} - L\,(a,\dot{a}) = 
  a\;
  \Bigl\{
   - \displaystyle\frac{3}{8\pi\,G}\:
    \bigl[\dot{a}^{2} + k \bigr] +
    a^{2}\,\rho(a)
  \Bigr\},
\end{array}
\label{eq.model.1.5}
\end{equation}
where $p$ is momentum conjugated to generalized coordinate $a$,
we find \cite{Maydanyuk.2011.EPJP}:
\begin{equation}
\begin{array}{cll}
  h\,(a,p_{a}) & = &
  -\:\displaystyle\frac{1}{a}\;
  \biggl\{
    \displaystyle\frac{2\pi\,G}{3}\: p_{a}^{2} +
    a^{2}\,\displaystyle\frac{3\,k}{8\pi\,G} -
    a^{4}\,\rho(a) \biggr\}.
\end{array}
\label{eq.model.1.6}
\end{equation}
We apply quantization and obtain the stationary Wheeler-De Witt equation
(for example, see~\cite{Vilenkin.1995}, (16)--(17) p.~4; see~\cite{Wheeler.1968,DeWitt.1967,Rubakov.2002.PRD}):
\begin{equation}
\begin{array}{lll}
\vspace{1mm}
  \biggl\{
    -\: \displaystyle\frac{\partial^{2}}{\partial a^{2}} +  \bar{V} (a)
  \biggr\}\; \varphi(a) = \bar{E}_{\rm rad}\; \varphi(a), & \\

\vspace{1.7mm}
  \bar{V} (a) = \displaystyle\frac{12}{8\pi\,G}\;
    \Bigl[
      \displaystyle\frac{3}{8\pi\,G}\: k\,a^{2} -
      \rho_{\rm gCg}(a)\, a^{4}
    \Bigr], & \\

  \bar{E}_{\rm rad} = \displaystyle\frac{3\, \rho_{\rm rad}}{2\pi\,G}.
\end{array}
\label{eq.model.1.7}
\end{equation}

\subsection{Motivations for non-stationary generalization of Wheeler-De Witt equation
\label{sec.model.2}}

Let us analyze how the velocity of the evolution of the universe can be defined in quantum approach. In classical mechanics the velocity of the particle is related to its momentum. In quantum mechanics, there is connection between the corresponding operators. According to basic positions of quantum mechanics, determination of the wave function $\Psi$ at some moment of time $t_{0}$, not only fully describes all quantum properties of the studied system at this time $t_{0}$, but also fully determines its evolution at all future times.
In other words, derivative $\partial \Psi / \partial t$ of the wave function over time at any given moment of time  $t_{0}$ is determined by this wave function $\Psi$ at $t_{0}$~\cite{Landau.v3.2004}. In particular, such an idea is used in the formulation of the non-stationary Schr\"{o}dinger equation in the non-relativistic quantum mechanics. However, this initial idea (followed from sense of the wave function in quantum physics) does not restrict us to use only the differential derivatives over time of the first order and forms of hamiltonian in the non-relativistic case.
It also covers possibility to involve the differential derivatives over time of higher orders and other time operators acting on the wave functions. If we want to use the principle of superposition, the relations between separated terms should be linear.
On this basis, instead of the stationary Wheeler-De Witt equation we shall introduce the following equation
(which we shall consider further up to terms with derivatives over time by the second order):
\begin{equation}
\begin{array}{ccc}
  i\,\hbar_{1}\: \displaystyle\frac{\partial \Psi}{\partial t} +
  \hbar_{2}\: \displaystyle\frac{\partial^{2} \Psi}{\partial t^{2}} +
  o\, \Bigl( \displaystyle\sum_{n=3} \hbar_{n}\: \displaystyle\frac{\partial^{n} \Psi}{\partial t^{n}} \Bigr) +
  \hat{\varepsilon} (t) \Psi =
  \hat{h}\: \Psi,
\end{array}
\label{eq.model.2.1}
\end{equation}
where $\hat{h}$ is some unknown operator,
$\hbar_{1}$, $\hbar_{2}$ and $\hbar_{n}$ are constants (which are supposed could be different, in general, and they determine relative contributions of the different non-stationary components),
$\hat{\varepsilon} (t)$ is operator describing possible interference between different components.

This equation is not direct continuation of the hamiltonian determined before, as action and following lagrangian and hamiltonian are determined in classical dynamics. Quantum physics should be not direct result of the defined before classical dynamics, but it covers this classical dynamics, including it as partial case. Moreover, in dependence on the chosen transition to the classical physics the different equations of motion and corresponding hamiltonians can be obtained, and classical characteristics can obtain different formulations (for example, tunneling phenomenon is strongly different in frameworks of non-relativistic Scgr\"{o}dinger equation and Dirac equation, that can be confirmed after analysis of the corresponding wave functions).
Quantum chromodynamics is particular example when quantum theory is not only be obtained via quantization of classical theory, but it has no any classical analogue.

Quantum physics provides more independent degrees of freedom of the studied system, than classical physics. Quantum physics works with measured observables (for example, components of electric and magnetic fields) and not measurable characteristics hidden for classical physics (for example, components of vector potential of electromagnetic field, where gauges are imposed further).
By such motivations it is fair to conclude that the introduced quantum equations should contradict to the existed classical equations of Einstein.
Use of the classical equations of cosmology as final criterium of validity of formulation of the quantum equations (in result of direct quantization) has no so proper basis (as it has smaller number of independent degrees of freedom).
Now, if we consider formulation of the non-stationary non-relativistic quantum mechanics \cite{Landau.v3.2004}, we shall see that it is not direct result of the chosen quantization procedure of classical mechanics. But, its stationary limits is related with classical mechanics (or wave optics).


For formulation of such a theory we shall choose a way of selection of the leading components,
which should describe the main quantum properties of the cosmological model.
In particular, if to consider in (\ref{eq.model.2.1}) the first item as leading term, then we obtain
\begin{equation}
\begin{array}{ccc}
  i\,\hbar_{1}\: \displaystyle\frac{\partial \Psi}{\partial t} =
  \hat{h}\: \Psi.
\end{array}
\label{eq.model.2.2}
\end{equation}
Note that we obtain this equation as not simple rewriting of form of the non-relativistic non-stationary Schr\"{o}dinger equation (as it is often supposed in literature). This equation can also include form of Dirac equation (with possibility to include interactions terms), and even Klein-Gordon equation, some forms of spinor equations, etc., that is dependent on the used hamiltonian (see \cite{Ahiezer.1981}).
In case of the leading second term the resulting equation continues line of Klein-Gordon equations.
However, this Klein-Gordon equation can be exactly presented in the form (\ref{eq.model.2.2}) (see \cite{Ahiezer.1981}, p.~9--10 for details). So, Eq.~(\ref{eq.model.2.2}) includes different formulations of Klein-Gordon equation and next developments of spinor physics.


As next step, before further study of all peculiarities which quantum equation~(\ref{eq.model.2.1}) gives us, it is better to see which results such a way can give us. In order to perform this, we shall restrict ourselves in this paper by consideration of eq.~(\ref{eq.model.2.2}) (further, we shall omit bottom index at constant $\hbar$).
Let us clarify what the operator $\hat{h}$ should correspond to. If to assume that the universe expands classically (i.e. with high degree of confidence we can neglect by the quantum properties) at large values of the scale factor $a$,
then the wave function can be written as
\begin{equation}
  \Psi = A\, e^{iS / \hbar},
\label{eq.model.2.3}
\end{equation}
where $S$ is action. Substituting this expression for the wave function into (\ref{eq.model.2.1}) and neglecting by change of amplitude $A$ over time,
we find:
\begin{equation}
\begin{array}{ccc}
  \displaystyle\frac{\partial \Psi}{\partial t} =
  \displaystyle\frac{i}{\hbar}\,
  \displaystyle\frac{\partial S}{\partial t}\,
  \Psi,
\end{array}
\label{eq.model.2.4}
\end{equation}
Comparing this expression with definition (\ref{eq.model.2.1}), we conclude that in the limiting case the operator $\hat{h}$ is reduced to simple multiplication on value of $-\partial S / \partial t$, i.e. Hamiltonian function.
Now we have introduced time into quantum equation, connecting it with Hamiltonian operator and taking into account that the asymptotic representation of the wave function is connected with action as (\ref{eq.model.2.2}).


\subsection{Operators of function of Hubble and velocity of expansion of the universe
\label{sec.model.3}}

As the non-stationary quantum equation has already been defined, now we can define operator of the velocity using general rule of differentiation of operators over time
(for example, see (9.2), (19.1) in \cite{Landau.v3.2004}).

\vspace{5mm}
\noindent
\textbf{Definition 1.} \emph{We define operator of the velocity as}
\begin{equation}
\begin{array}{ccc}
  \hat{\dot{a}} = \displaystyle\frac{i}{\hbar}\; (\hat{h}\, a - a\, \hat{h}).
\end{array}
\label{eq.model.3.4}
\end{equation}
%
%
\noindent
Here, we use requirement (as in standard quantum mechanics):
\begin{equation}
  \displaystyle\frac{\partial}{\partial t} \displaystyle\int |\Psi(a, t)|^{2}\; da = 0.
\label{eq.model.3.5}
\end{equation}
Substituting hamiltonian, we find\footnote{Note that the operator of velocity~(\ref{eq.model.3.6}) is different from the standard definition of the operator of velocity in the non-relativistic quantum mechanics (for example, see (19.1) in \cite{Landau.v3.2004}),
as it is introduced on the basis of the non-linear hamiltonian~(\ref{eq.model.1.6}). On such a reason, knowledge of the differential equation (\ref{eq.model.5.2}) and wave function is not enough for full determination of
 velocity (in contrast to the standard non-relativistic quantum mechanics).}:
\begin{equation}
\begin{array}{lll}
  \hat{\dot{a}} =
  \displaystyle\frac{i}{\hbar}\,
  \displaystyle\frac{8\pi G}{6\,a }\;
  \displaystyle\frac{\partial}{\partial a}.
\end{array}
\label{eq.model.3.6}
\end{equation}

\vspace{5mm}
\noindent
\textbf{Definition 2.} \emph{We define operator of function of Hubble as}
\begin{equation}
  \hat{H}\,(a) = \displaystyle\frac{1}{a}\,\hat{\dot{a}}.
\label{eq.model.3.7}
\end{equation}
%
%
\noindent
According to this definition, we shall consider the parameter of Hubble at the given scale factor $a_{0}$, as action of certain operator $\hat{H}$ on the wave function at $a_{0}$. The wave function is not eigenfunction of operators of the velocity and the function of Hubble, as there are no any constant eigenvalues for these operators.
So, action of these operators on the wave function can be written as
\begin{equation}
\begin{array}{ll}
  \hat{\dot{a}}\: \Psi\,(a) = v(a)\, \Psi\,(a), &
  \hat{H}\,(a)\, \Psi\,(a) = H\,(a)\, \Psi\,(a).
\end{array}
\label{eq.model.3.8}
\end{equation}
Here, $v(a)$ and $H\,(a)$ are some functions changed in depending on $a$. Near to arbitrarily chosen value $a_{0}$ these functions tend to certain well-defined fixed values, which can be locally considered as eigenvalues of operators of the velocity and the function of Hubble at $a_{0}$. Thus, on the basis of the functions $v(a)$ and $H\,(a)$ we shall understand the velocity and the parameter of Hubble in the quantum approach.
For practical calculations, one can obtain these functions as
\begin{equation}
\begin{array}{lll}
  v\,(a) =
  \displaystyle\frac{\hat{\dot{a}}\; \Psi\,(a)}{\Psi\,(a)}, &
  H\,(a) =
  \displaystyle\frac{\hat{H}\; \Psi\,(a)}{\Psi\,(a)} =
  \displaystyle\frac{v\,(a)}{a}.
\end{array}
\label{eq.model.3.9}
\end{equation}

\subsection{Quantum definition of duration of existence of universe
\label{sec.model.4}}

As the function of Hubble has been defined, as next step we can define duration of existence of the universe. We shall be interesting in such a characteristic which is dependent on the scale factor $a$: this should allow to see clearly behavior of expansion (evolution) and compare such a dynamic for different model scenarios. So, we introduce the following characteristic:
\begin{equation}
\begin{array}{lll}
  t\,(a) =
  \displaystyle\int\limits_{a_{\rm min}}^{a}
  \displaystyle\frac{1} {\tilde{a}\, H\,(\tilde{a})}\; d\tilde{a}.
\end{array}
\label{eq.model.4.1}
\end{equation}
One can see that such a definition is analog of the classical definition of the universe age given in classical cosmology (for example, see~(36), p.~11 in~\cite{Trodden.TASI-2003}).

\subsection{Rescaling
\label{sec.model.5}}

Inside region $0<a<100$ potentials of the considered models above achieve essential values (that can cause serious difficulties in practical calculations and further analysis). But, they can be decreased via renormalization of the corresponding equations (\ref{eq.model.1.7}) (we shall call such a procedure as \emph{rescaling}).
By such a reason, let us pass to a new variable:
\begin{equation}
\begin{array}{cc}
  a_{\rm new} = \nu\, a_{\rm old}, &
  \hspace{5mm}
  \nu = \sqrt{\displaystyle\frac{12}{8\pi\,G}}.
\end{array}
\label{eq.model.5.1}
\end{equation}
Now from (\ref{eq.model.1.7}) we obtain new equations for determination of the wave function:
\begin{equation}
\begin{array}{lll}
\vspace{1.4mm}
  \biggl\{
    -\: \displaystyle\frac{\partial^{2}}{\partial a_{\rm new}^{2}} +
    V \Bigl(\displaystyle\frac{a_{\rm new}}{\nu}\Bigr)
  \biggr\}\; \varphi(a_{\rm new}) = E_{\rm rad}\; \varphi(a_{\rm new}), & \\

\vspace{2mm}
  V (a) = \displaystyle\frac{1}{8\pi\,G}\;
    \Bigl[
      \displaystyle\frac{3}{8\pi\,G}\: k\,a^{2} -
      \rho_{\rm gCg}(a)\, a^{4}
    \Bigr], & \\

  E_{\rm rad} = \displaystyle\frac{\rho_{\rm rad}}{8\pi\,G}.
\end{array}
\label{eq.model.5.2}
\end{equation}
For operator of velocity (\ref{eq.model.3.6}), determined via new variable, we obtain:
\begin{equation}
\begin{array}{lll}
  \hat{\dot{a}} =
  \displaystyle\frac{i}{\hbar}\,
  \displaystyle\frac{8\pi G\, \nu^{2}}{6\,a_{\rm new}}\;
  \displaystyle\frac{\partial}{\partial a_{\rm new}}.
\end{array}
\label{eq.model.5.3}
\end{equation}

\section{Analysis
\label{sec.results.2}}

\subsection{Tunneling deeply under the barrier
\label{sec.results.2.1}}

Now we shall demonstrate how the formulation of quantum dynamics works on the cosmological model above.
We shall use eq.~(\ref{eq.model.5.2}) at the energy density defined in (\ref{eq.model.1.2}) for the closed Universe ($k=1$).
Some estimations of rates were performed on the basis of analysis of wave packet tunneling through the potential barrier in the standard formulation of Chaplygin gas~\cite{Monerat.2007.PRD}.
In order to give other researchers basis for comparison and analysis, we shall choose $\bar{A}=0.001$, $\bar{B}=0.001$, according to that paper.

In the beginning let us consider tunneling deeply under the barrier where we choose energy of radiation
of $E_{\rm rad}=100$.
In particular, such energy region is hidden for the semiclassical approaches.
\begin{figure}[htbp]
\centerline{\includegraphics[width=92mm]{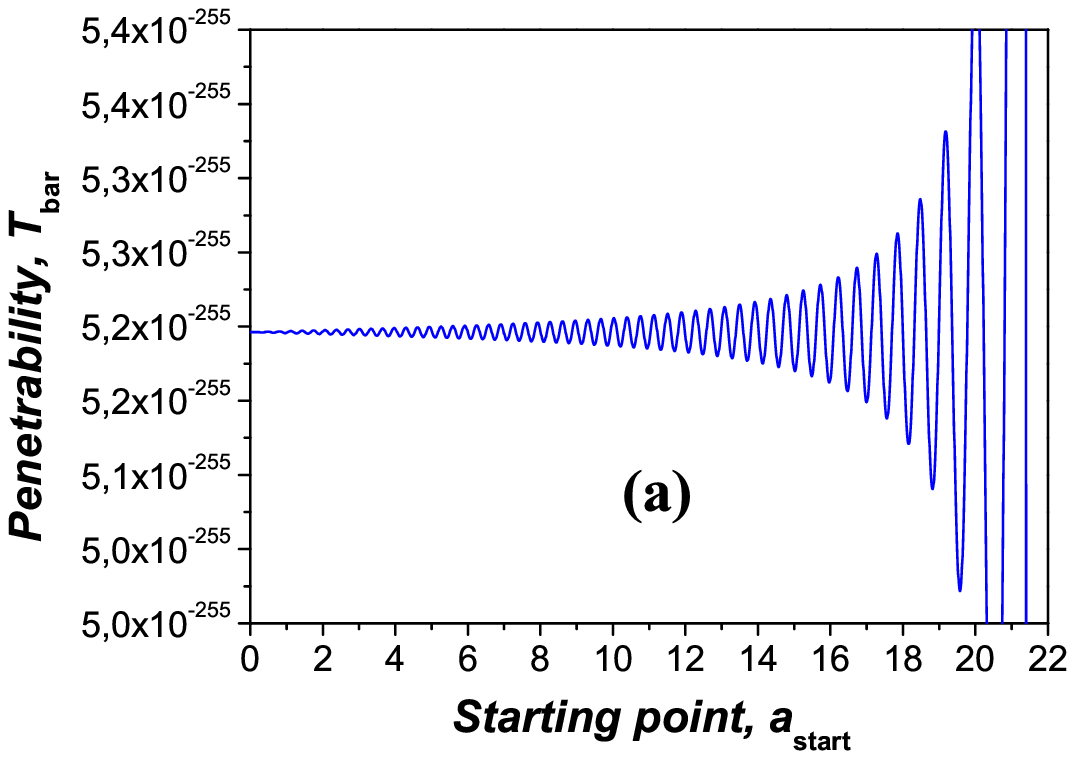}}
\vspace{-7mm}
\centerline{\includegraphics[width=92mm]{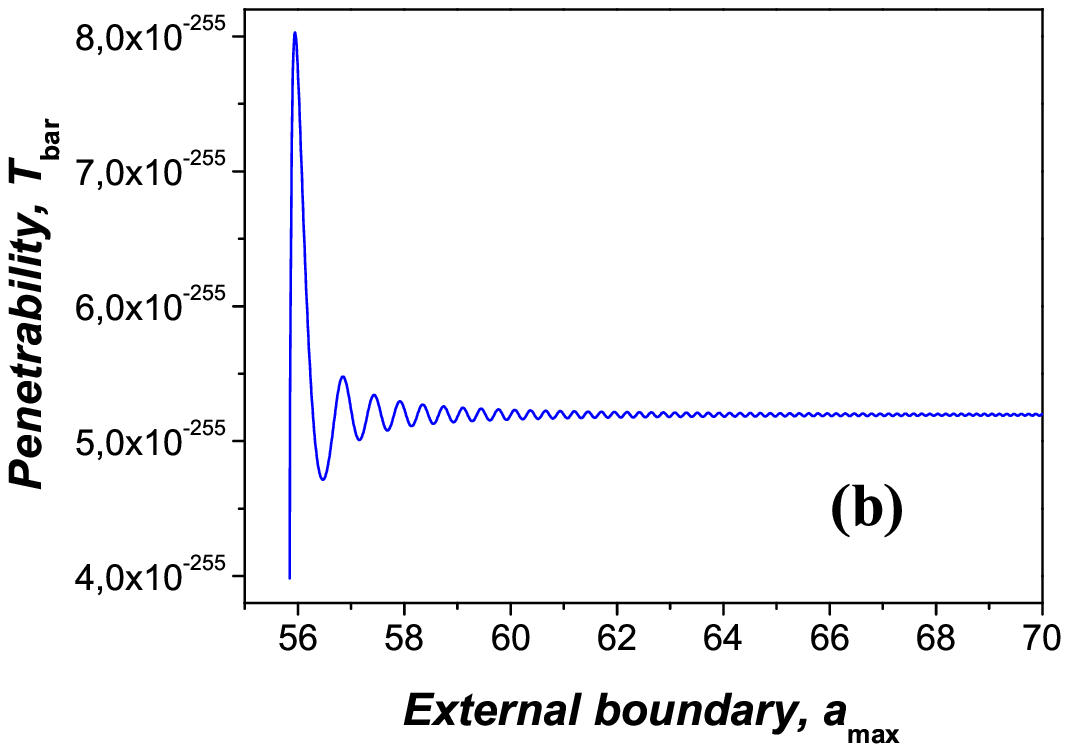}}
\vspace{0mm}
\caption{\small 
The penetrability of the barrier for model~\cite{Maydanyuk.2011.EPJP} at $E_{\rm rad}=100$ (parameters of calculation: 10000 intervals at $a_{\rm max}=100$):
(a) dependence of penetrability on the starting point $a_{\rm start}$ in region from zero up to internal turning point, at fixed $a_{\rm max}=100$:
at increasing of $a_{\rm start}$ the penetrability oscillates, maxima are increased and minima are decreased,
(b) dependence of the penetrability on external boundary $a_{\rm max}$ at fixed $a_{\rm start}=0.1$ (corresponding to coordinate of minimum of the internal well before the barrier): one can see that at increasing of $a_{\rm max}$ the calculated penetrability tends to some definite limit value, which we chose for further calculations and analysis.
\label{fig.2}}
\end{figure}
The calculated penetrability in dependence on the starting point $a_{\rm start}$ has oscillating behavior, maxima are slowly increased and minima are decreased at tending of the starting point to internal turning point (see Fig.~\ref{fig.2}~(a)).
At increasing of the external boundary $a_{\rm max}$ (starting from the external turning point) the penetrability tends to some definite value (see Fig.~\ref{fig.2}~(b)).
From such calculations we conclude:
(1) determination of the penetrability on the basis of shape of the barrier inside tunneling region only (used in the semiclassical approaches) is far from the penetrability calculated after taking into account external tail and internal well,
(2) in spite of sharp decreasing of the potential in the external region (after external turning point) at increasing of the external boundary $a_{\rm max}$ calculations of the penetrability are convergent (that allows us to say about reliable values of the penetrability for such potentials).

\begin{figure}[htbp]
\centerline{\includegraphics[width=92mm]{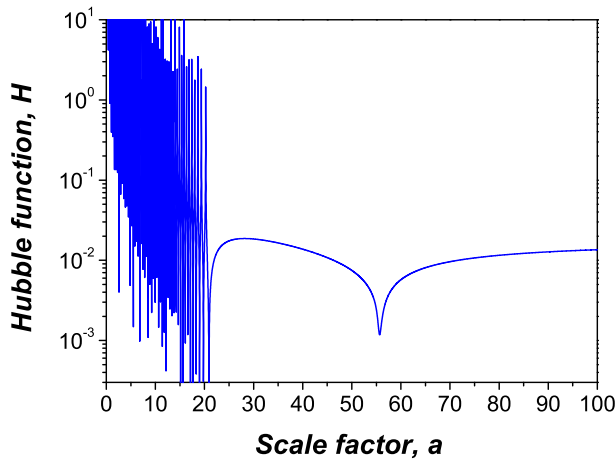}}
\caption{\small 
The modulus of the function of Hubble in dependence on scale factor $a$ at $E_{\rm rad}=100$ (calculation parameters: starting point $a_{\rm start}=0.1$; 10000 intervals at $a_{\rm max}=100$):
whole range of variable $a$ can be separated on 3 regions: in the internal region (at $a<21$) presence of chaotic peaks and minima is observed, in the tunneling region (at $21<a<55.6$) the function is smooth and has no any oscillation, in the external region it increases monotonously (slowly transforming to linear dependence).
\label{fig.3}}
\end{figure}

\begin{figure}[htbp]
\centerline{\includegraphics[width=92mm]{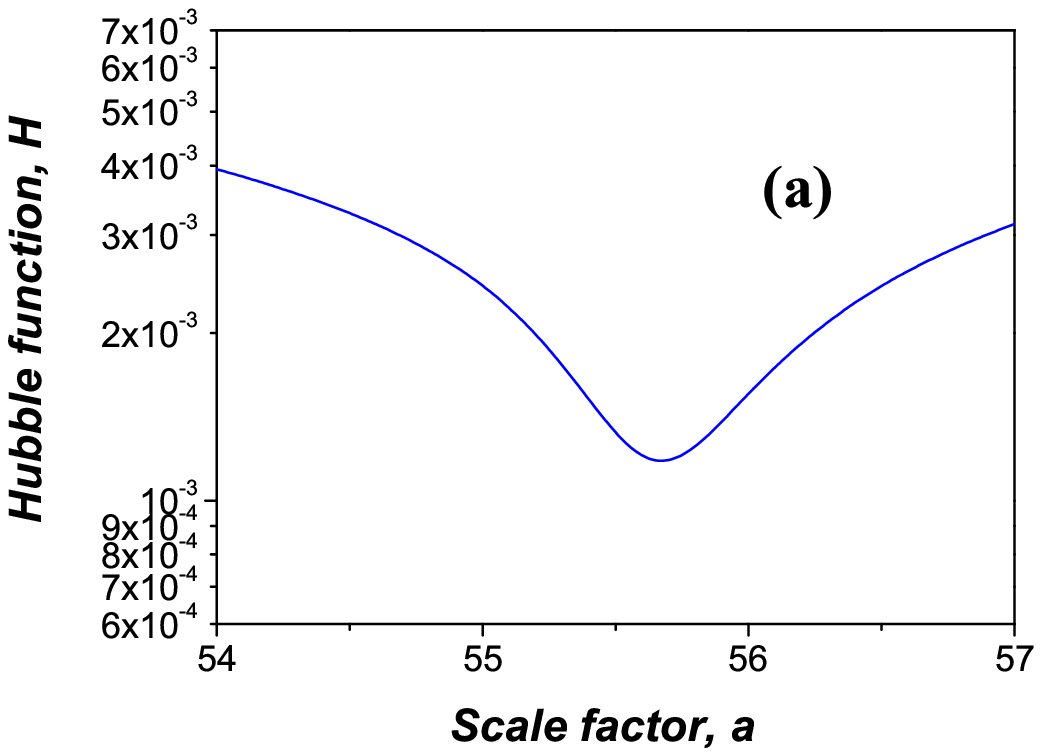}}
\vspace{-5mm}
\centerline{\includegraphics[width=92mm]{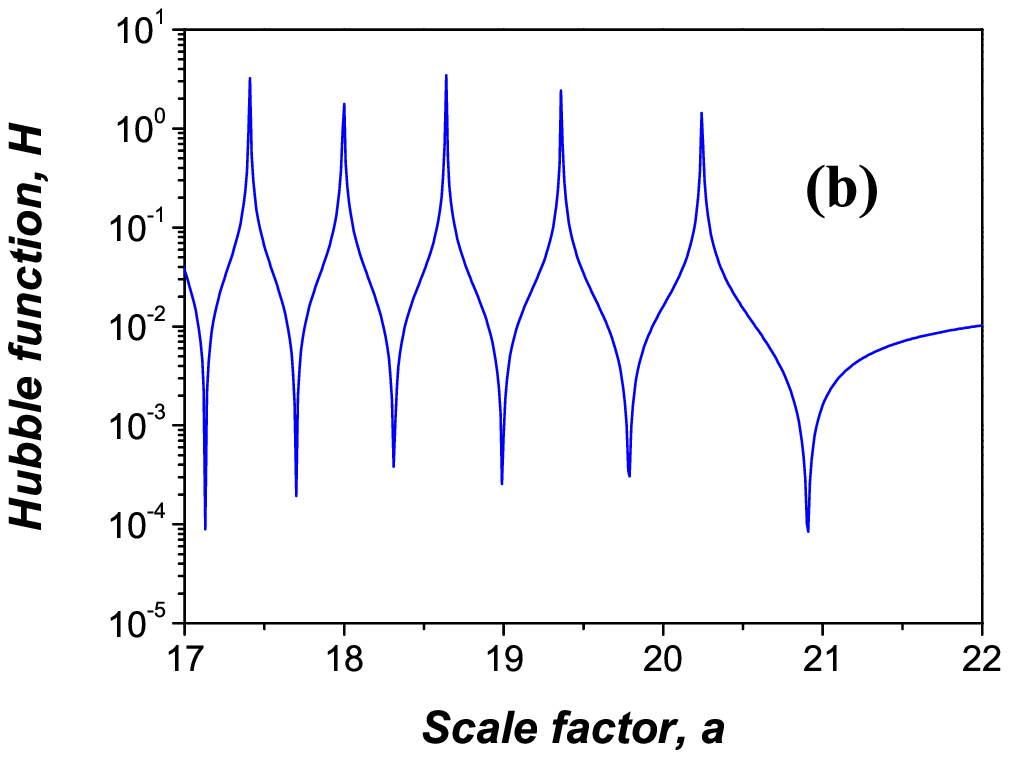}}
\vspace{0mm}
\caption{\small 
The modulus of the function of Hubble in dependence on scale factor $a$ at $E_{\rm rad}=100$ (calculation parameters: starting point $a_{\rm start}=0.1$; 10000 intervals at $a_{\rm max}=100$):
(a) at transition from the tunneling region to external one smooth minimum of this function is observed, corresponding to external turning point,
(b) location of minima and peaks at small $a$ is similar (the last minimum corresponds to the internal turning point).
\label{fig.4}}
\end{figure}

Now we shall analyze the function of Hubble defined by formula~(\ref{eq.model.3.9}).
In Fig.~\ref{fig.3} a general picture how the modulus of this function is changed in dependence on scale factor $a$ is shown.
At small $a$ (close to $a=21$) sharp chaotic peaks are observed, then this function has one clear maximum and minimum, at finishing it increases monotonously. In the first consideration, such a behavior of the function of Hubble (especially at small $a$) looks to be enough strange.
But after increasing insight of the last right minimum at $a=55.6$ one can see (see Fig.~\ref{fig.4}~(a)) that it is stable and is not equal to zero (that confirms convergence of calculations), corresponding to the external turning point --- i.e. it separates the tunneling region from the external region. In the external region the modulus of the function of Hubble increases monotonously, slowly transforming to linear dependence.
Inside the region $21<a<55.6$ the modulus of the function of Hubble has no any oscillation --- this is the tunneling region, which is finished by convergent minima at both sides.
Detailed analysis of the internal region (at $a<21$) shows that location of minima and peaks here is similar (see Fig.~\ref{fig.4}~(b)), the last minimum corresponds to the internal turning point.

\begin{figure}[htbp]
\centerline{\includegraphics[width=92mm]{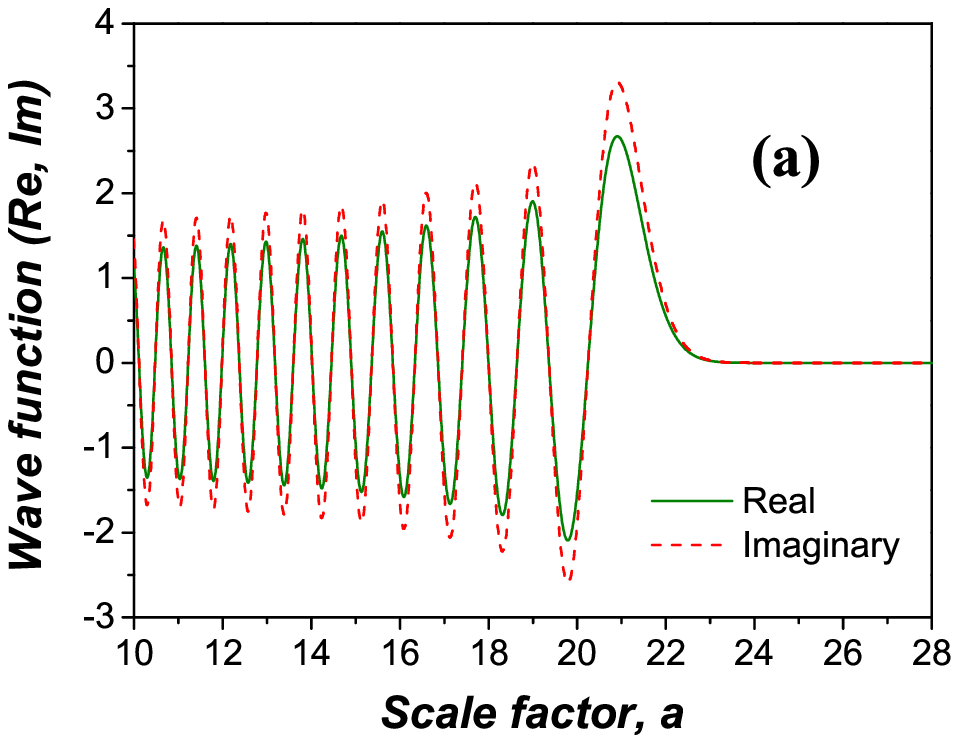}}
\vspace{-8mm}
\centerline{\includegraphics[width=92mm]{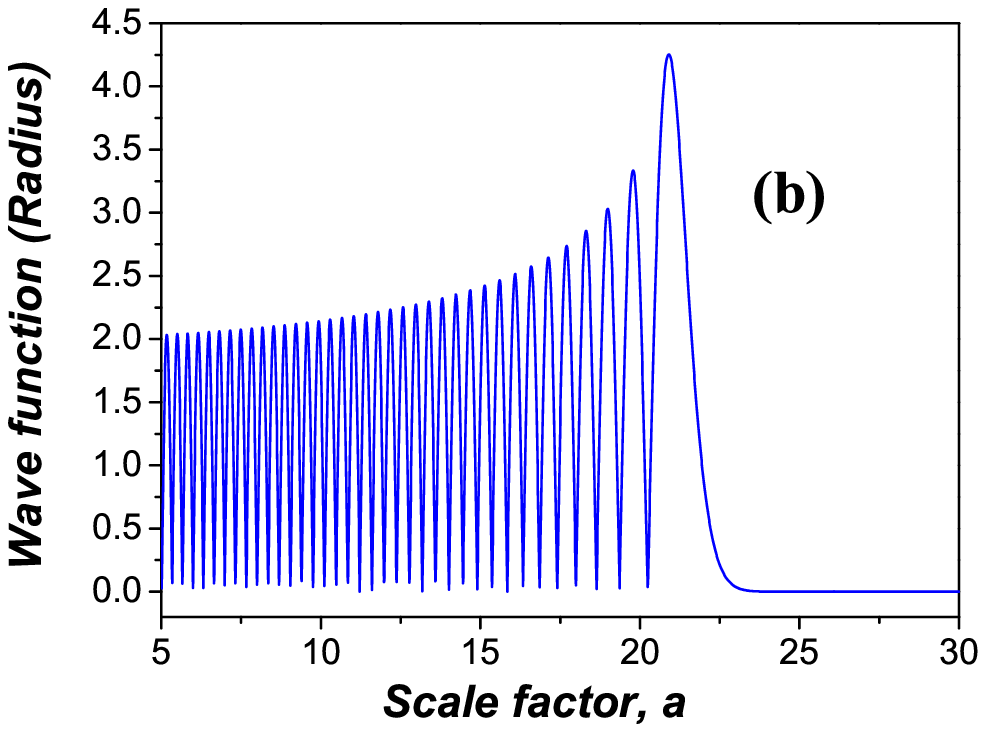}}
\vspace{-2mm}
\caption{\small 
The wave function in dependence on the scale factor $a$ at $E_{\rm rad} = 100$ (calculation parameters: the starting point $a_{\rm start} = 0.1$; 10000 intervals at $a_{\rm max} = 100$)
(a) in the internal region (at $a<21$) accelerated growth of the maxima of modulus of the wave function (with minima very closed to zero) up to the internal turning point is observed, in the tunneling and external regions (at $21<a$) this modulus decreases without oscillations,
(b) in the internal region the real and imaginary parts of the wave functions oscillate almost simultaneously, while in the tunneling and external regions they extremely fall down
[consideration of the wave function in the logarithmic scale clearly demonstrates its non-zero values in the tunneling and the external regions].
\label{fig.5}}
\end{figure}

The following question can be appeared: whether peaks are finite at small $a$ or we are dealing with divergencies in calculations? Let us consider formula (\ref{eq.model.3.9}) for the function of Hubble: one can see that maximal values should be caused by practically zero values of the wave function, which is in the denominator.
The wave function calculated for this process is shown in Fig.~\ref{fig.5}.
In the first figure (a) the modulus of the wave function is shown: indeed, with clearly determined finite maxima (which should definitely give not zero minima of the function of Hubble) the sharp minima are seen tending to zero --- it explains presence of the sharp peaks in the Hubble function in Fig.~\ref{fig.2}. This could be explained by almost simultaneous zeroing of the real and imaginary parts of the wave function. However, this is strange as we have non-zero complex wave function (as it defines non-zero constant flux directed outside inside entire region of variable $a$), so zeros of its real and imaginary parts should not be coincide anywhere.
In the next figure (b) the real and imaginary parts of the wave function are shown. Here, one can see clear stable curves that demonstrates convergence and stability of calculations, and we need to understand the result. However, the curves behave almost similarly: their maxima, minima and zeroes are located at close coordinates. This situation is similar to the behavior of the wave function of the bound state for a particle inside a potential well (with infinitely high boundaries). Indeed, at the chosen energy the penetrability is very small ($ T_{\rm bar} \sim 10^{-255}$) and output is extremely small. By other words, we are dealing with the quasi-stationary state with extremely small output outside, which is very close to stationary one, practically.




\subsection{Tunneling near the barrier maximum and above-barrier propagation processes
\label{sec.results.2.2}}

Now let us consider a case where the tunneling occurs near the barrier maximum. In this case, we can use our previous analysis in~\cite{Maydanyuk.2011.EPJP} and choose $E_{\rm rad} = 220$. So, at start in point $a_{\rm start} = 0.1$ we obtain $T_{\rm bar} = 1.52129237224042 \cdot 10^{-7}$, $R_{\rm bar} = 0.999999847870763$ and the condition $T_{\rm bar} + R_{\rm bar} = 1$ holds up to 14 digits\footnote{Note that the semiclassical methods usually do not give the reflection coefficient $R_{\rm bar}$ and the mentioned test is not applied. However, in discussions on comparison between fully quantum calculations and semiclassical ones this important point is usually ignored, with assumptions on advantage of the semiclassical apparatus without alternatives.}.

\begin{figure}[htbp]
\vspace{-8mm}
\centerline{\includegraphics[width=92mm]{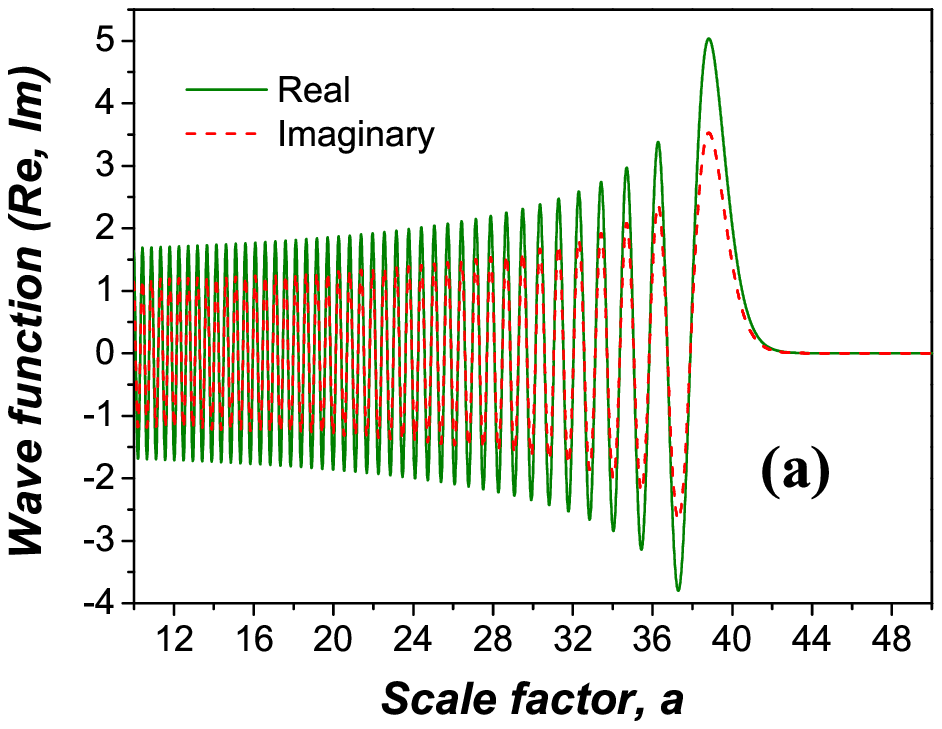}}
\vspace{-9mm}
\centerline{\includegraphics[width=92mm]{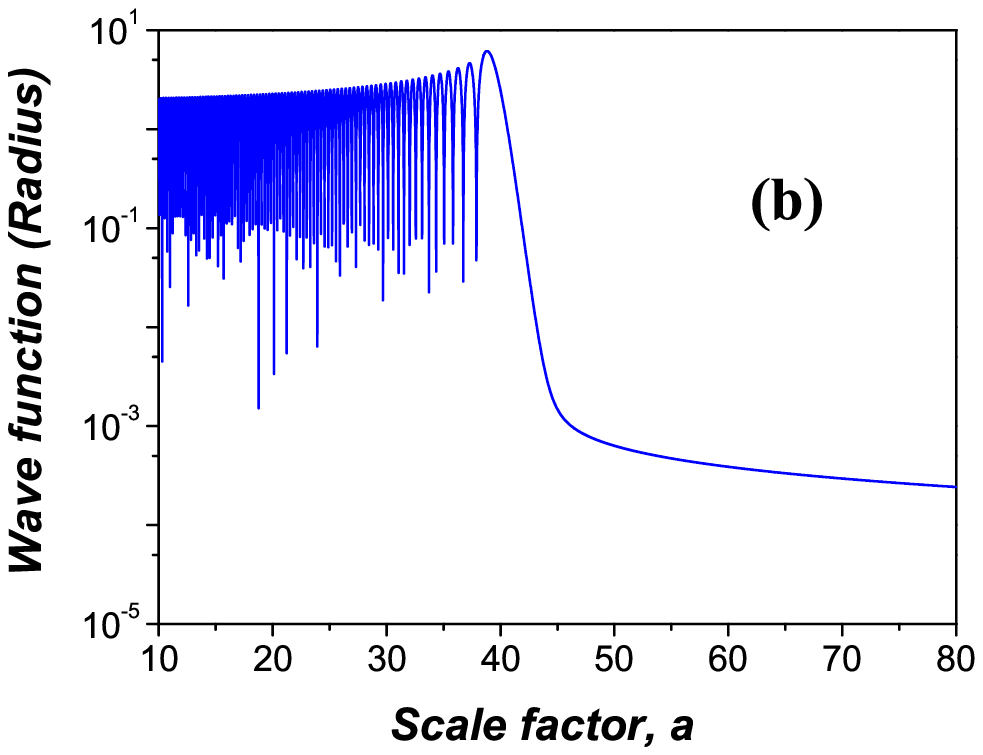}}
\vspace{-4mm}
\caption{\small 
The wave function in dependence on the scale factor $a$ at $E_{\rm rad} = 220$ (calculation parameters: the starting point $a_{\rm start} = 0.1$, 10000 intervals at $a_{\rm max} = 100$):
(a) in the internal region number of oscillations of the wave function is much larger and its maxima increase stronger in comparison with results at $E_{\rm rad} = 100$ (see Fig.~\ref{fig.4}~(b)), that is explained by increasing of the energy $E_{\rm rad}$ and enlarging of the internal well region;
(b) in the tunneling and external regions the modulus of the wave function is changed essentially weaker in comparison with results at $E_{\rm rad} = 100$ (see Fig.~\ref{fig.4}~(c)), that indicates on essential oncoming of energy $E_{\rm rad}$ to the barrier maximum.
\label{fig.6}}
\end{figure}

\begin{figure}[htbp]
\vspace{-9mm}
\centerline{\includegraphics[width=92mm]{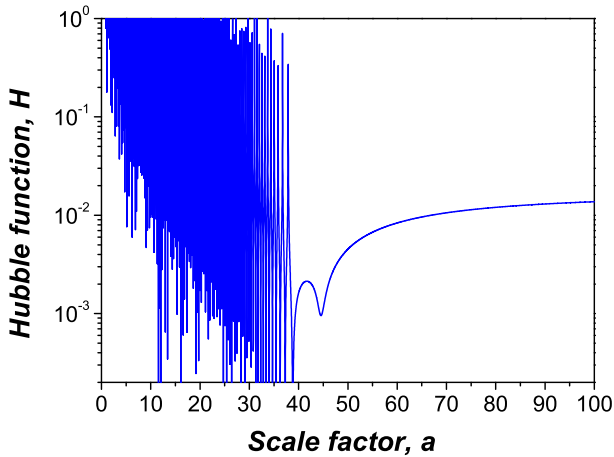}}
\vspace{-1mm}
\caption{\small 
The function of Hubble in dependence on the scale factor $a$ at $E_{\rm rad} = 220$ (calculation parameters: the starting point $a_{\rm start} = 0.1$, 10000 intervals at $a_{\rm max} = 100$):
the function of Hubble behaves similarly to the case at $E_{\rm rad} = 100$ (see Fig.~\ref{fig.3}), but here the tunneling region is much smaller and the function of Hubble in it is smaller.
\label{fig.7}}
\end{figure}

Our calculations show that the penetrability in dependence on the starting point $a_{\rm start}$ and on the external boundary $a_{\rm max}$ behaves like to the case studied above at $E_{\rm rad} = 100$ (also see~\cite{Maydanyuk.2011.EPJP}).
Placing the starting point in the minimum of the internal well, we calculate the wave function (see Fig.~\ref{fig.6}) and the function of Hubble (see Fig.~\ref{fig.7}).
From these figures it is clear that these functions behave as obtained above at $E_{\rm rad} = 100$. However, in this case we observe:
(1) in the internal region number of oscillations of the wave function is essentially larger and the difference between its maxima is reinforced (see Fig.~\ref{fig.6}~(a) in comparison with Fig.~\ref{fig.5}~(a)), caused by enlarging of the internal region with shift of the internal turning point to the right;
(2) in the tunneling and external regions the difference between the maxima and minima of modulus of the wave function is much smaller (see Fig.~\ref{fig.6}~(b) in comparison with Fig.~\ref{fig.5}~(b)), that indicates on essential oncoming of the energy $E_{\rm rad}$ to the barrier maximum and much stronger outgoing flux through the barrier outside;
(3) the tunneling region is less, which can be easily found by typical behavior of the function of Hubble (see Fig.~\ref{fig.7}).

\begin{figure}[htbp]
\vspace{-9mm}
\centerline{\includegraphics[width=92mm]{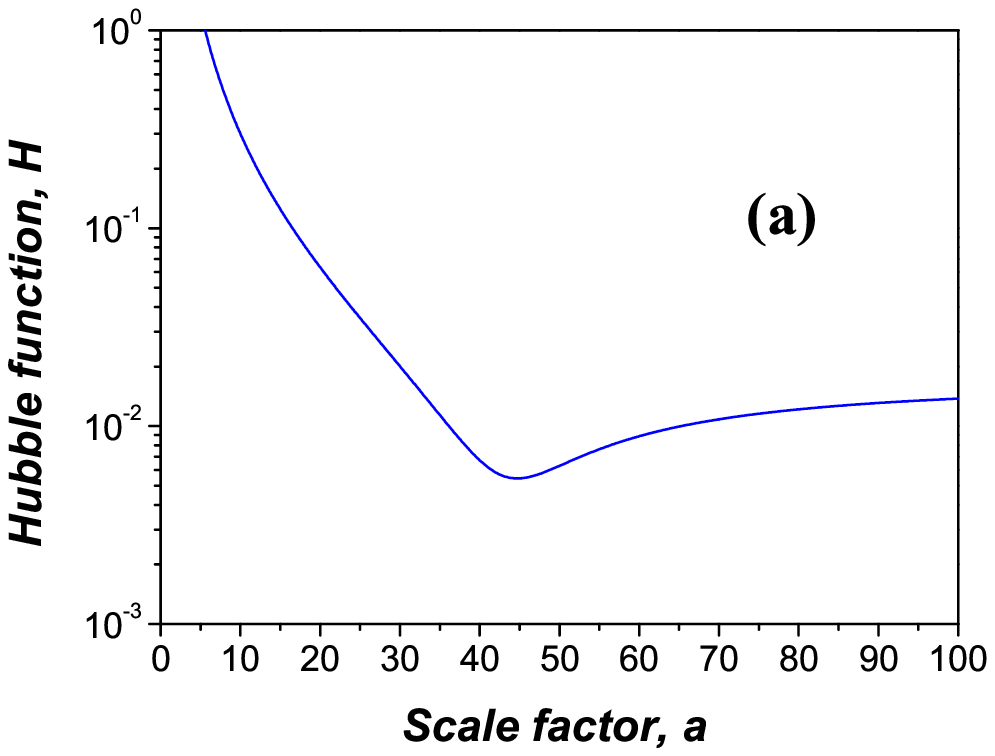}}
\vspace{-10mm}
\centerline{\includegraphics[width=92mm]{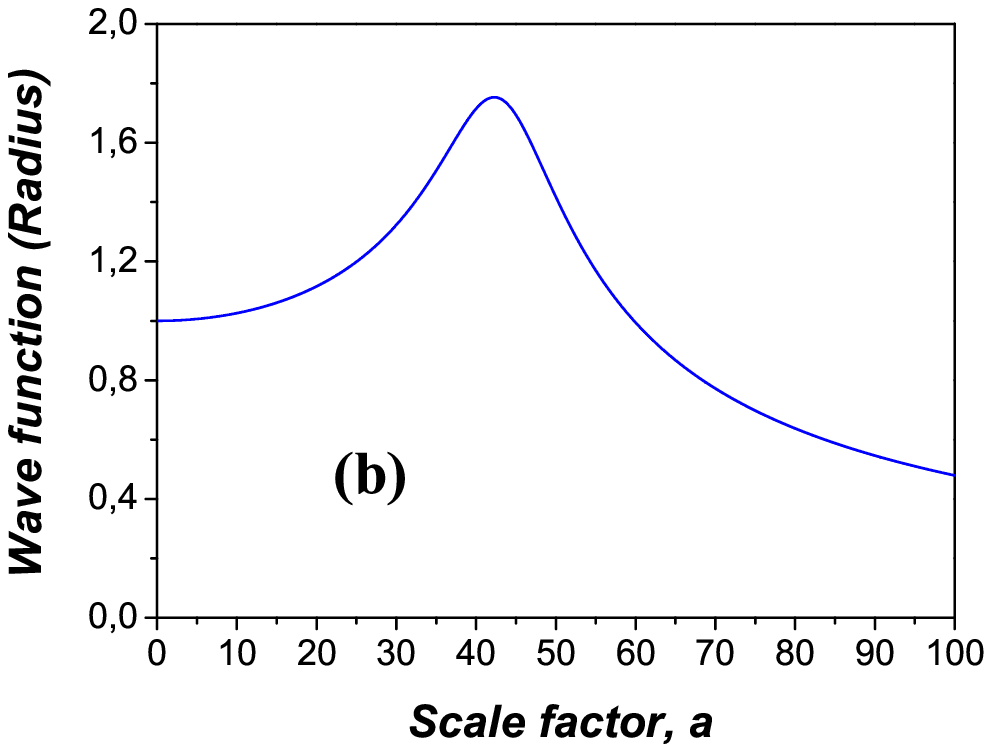}}
\vspace{-2mm}
\caption{\small 
The function of Hubble and wave function in dependence on the scale factor $a$ at $E_{\rm rad} = 250$ (calculation parameters: the starting point $a_{\rm start} = 0.1$, 10000 intervals at $a_{\rm max} = 100$):
(a) the function of Hubble has a smooth form: in the internal region the sharp peaks and minima (typical for above-barrier energies and reflecting bound state) are disappeared completely, in the middle region there is only one smooth minimum in the coordinate of the barrier maximum (tunneling maximum is disappeared as shown in Figs.~\ref{fig.4}~(a) and \ref{fig.7}), in the external region the accelerated increase of the function of Hubble remains
(as in Figs.~\ref{fig.4}~(a) and \ref{fig.7});
(b) modulus of the wave function has maximum in coordinate of the barrier maximum.
\label{fig.8}}
\end{figure}

Now let us find out how these pictures will be changed, if we increase the energy $E_{\rm rad}$ above the barrier.
Results such calculations are shown in Fig.~\ref{fig.8}, where we have chosen $E_{\rm rad} = 250$.
In the first figure (a) the function of Hubble is shown. One can see that
(1) In the internal region the sharp peaks and minima have disappeared completely,
(2) In the middle region the behavior of this function, typical for the tunneling region, has disappeared also, and instead there is a smooth minimum in the coordinate of the potential barrier maximum,
(3) In the external region the monotonic increasing of the function of Hubble remains, slowly becoming to linear dependence.
The modulus of the wave function is shown in next figure~(b). In this case, we see oscillatory behavior, typical for the above-barrier energies. However, its maximum is located at coordinate of the barrier maximum,
with a monotonic decrease to both sides.

\begin{figure}[htbp]
\centerline{\includegraphics[width=92mm]{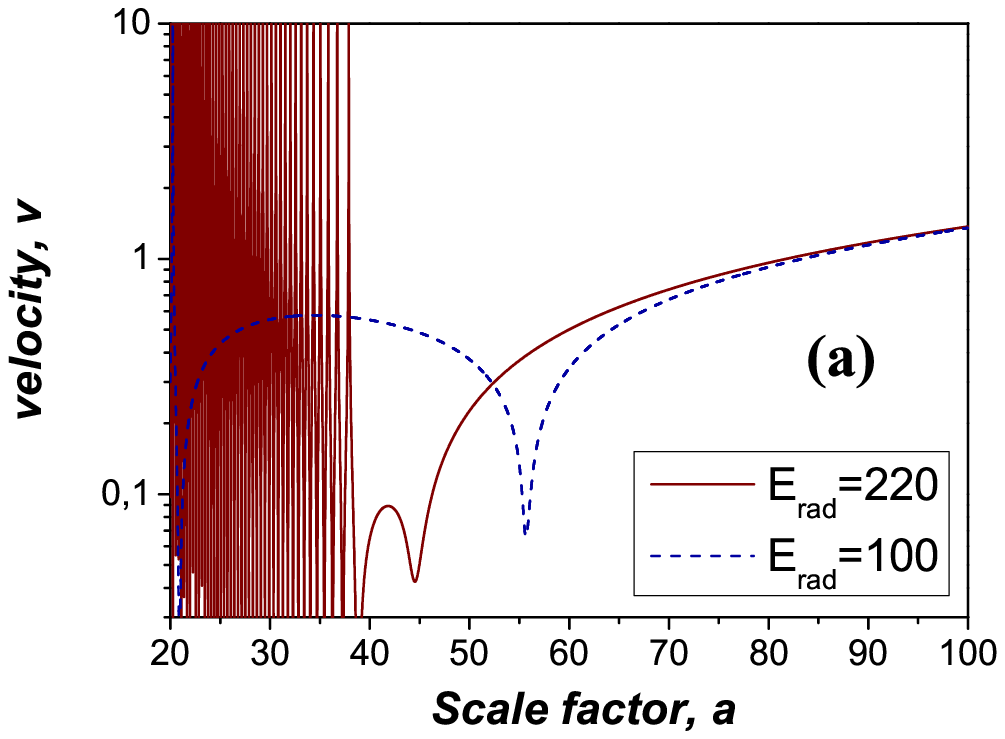}}
\vspace{-10mm}
\centerline{\includegraphics[width=92mm]{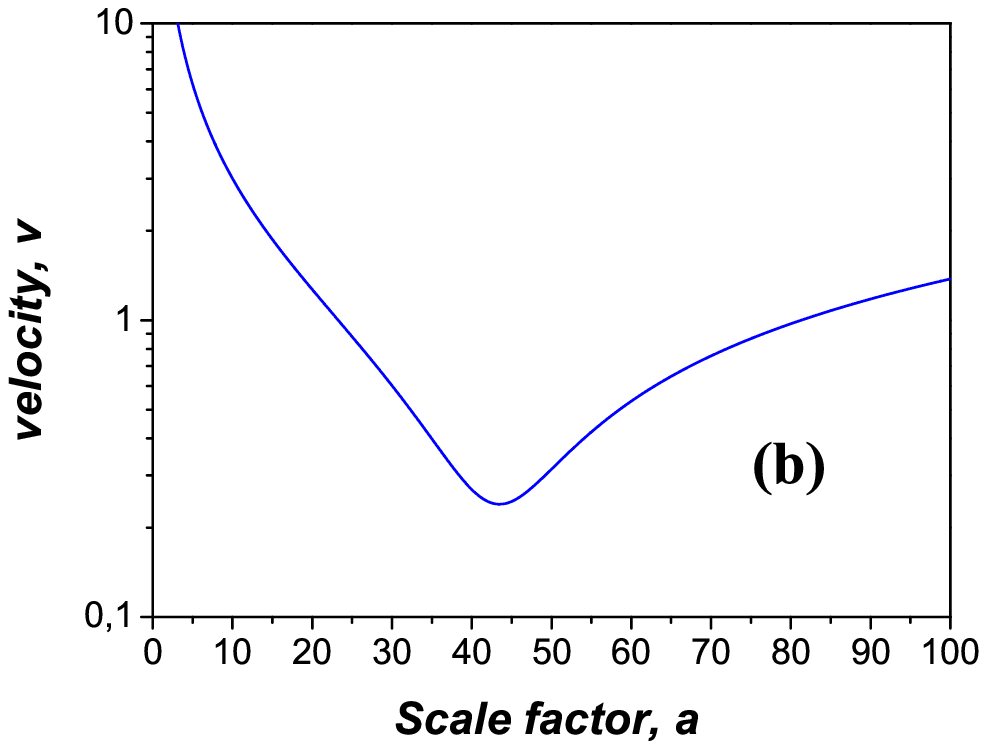}}
\vspace{-5mm}
\caption{\small 
The velocity of expansion of universe in dependence on the scale factor $a$ (calculation parameters: the starting point $a_{\rm start} = 0.1$, 10000 intervals at $a_{\rm max} = 100$):
(a) in the internal region at sub-barrier energies there are sharp peaks and minima, in the tunneling region the velocity is smooth, in the external region velocities at different energies tend to the same limit,
(b) for the above-barrier energy $E_{\rm rad} = 250$ sharp peaks and minima are disappeared, hump of tunneling (previously observed for the sub-barrier energies) transforms to smooth minimum in coordinate, corresponding to the potential barrier maximum.
\label{fig.9}}
\end{figure}

\begin{figure}[htbp]
\centerline{\includegraphics[width=92mm]{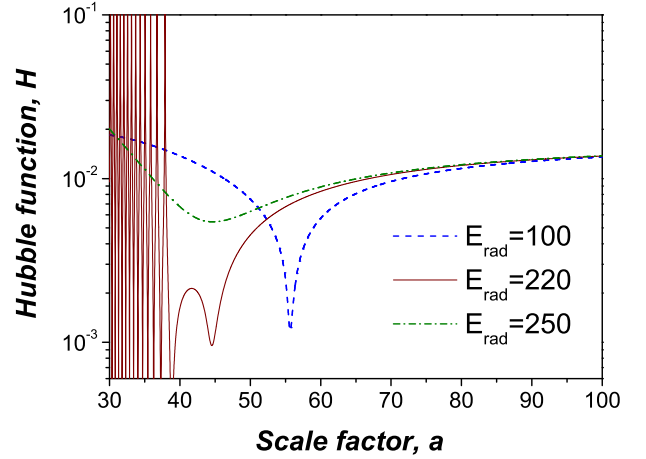}}
\vspace{-5mm}
\caption{\small 
The functions of Hubble at different energies $E_{\rm rad}$ tend to the same limit at large $a$
(calculation parameters: the starting point $a_{\rm start} = 0.1$, 10000 intervals at $a_{\rm max} = 100$).
\label{fig.10}}
\end{figure}

Now let us analyze, how velocity defined by the formula (\ref{eq.model.3.9}) behaves in each case.
Such calculations are presented in Fig.~\ref{fig.9}.
In general, the behavior of the velocity looks like the function of Hubble. From the figures it is clear that if in the internal region there are sharp peaks and minima at the sub-barrier energies (see figure~(a)), then they disappear at increasing of the energy $E_{\rm rad}$ above the barrier (see figure~(b)). However, after detailed analysis one can see the oscillations with a general tendency to decreasing (and decreasing amplitude with increasing $E_{\rm rad}$). They can be explained by wave nature, which quantum-mechanical treatment of the process gives us. They are the previously considered sharp peaks and minima at sub-barrier energies, i.e. now picture of sub-barrier and above-barrier processes becomes unite.
Fig.~\ref{fig.10} shows that the function of Hubble at different energies $E_{\rm rad}$ tend to the same limit behavior at large $a$.

So, there is the following picture at the above-barrier energy $E_{\rm rad} = 250$. We have no any classically forbidden region inside whole area of the scale factor $a$. Hence it would seem that this situation forbids to consider formation of the universe, and we would reject this case. However, we see that for values of $a$ less than coordinate of the barrier maximum (we denote it as $a_{\rm bar}$) the velocity gradually decreases to minimum with increasing of $a$. In this case, the modulus of the wave function increases up to maximum: it points to a gradual increase of probability of appearance of the universe, with maximum at $a_{\rm bar}$. If to take the spherically symmetric picture of the extension into account (where the density of matter in filled volume should not increase and it can be associated with the probability of appearance of the the universe), then such an increase of the probability is contrary to natural expansion of the universe in classical treatment. So, this situation is more suitable to description of formation of the universe with classical space-time, with its birth at $ a_{\rm bar}$. Starting from $a_{\rm bar}$, the expansion of space-time becomes classic. Before to this coordinate, the universe is formed with a gradual damping expansion, at maximum in point $a_{\rm bar}$ this extension is practically stops for a certain period, then further expansion begins with the acceleration. This logic points us to competence of quantum description of formation of the universe at the above-barrier energies (when there is no tunneling).

Also another property is found:
\emph{All velocities at different energies $E_{\rm rad}$ tend to the same limit with increasing $a$ in the external region: i.e. this model gives the same dynamics of the accelerated expansion of the universe for later times with completely different scenarios of its evolution in the first stage.}

\subsection{Time of evolution of the universe
\label{sec.results.3}}

Concluding such results, we see that the introduced above definitions for operators of the velocity of expansion of the universe and the function of Hubble give a clear basis for analysis of the dynamical evolution in its fully quantum consideration, supported by stable calculations. Now, if we hide such characteristics, then some calculations of changed shape of the wave packet will remain only for quantum description (estimation) of dynamics of evolution of the universe (see \cite{AcacioDeBarros.2007.PRD,Monerat.2007.PRD,Oliveira-Neto.2011.IJMPCS}). But this way is essentially more complicated and strongly higher unstable. It is very difficult to obtain proper results for not large values of the scale factors. And, practically, sense of the velocity and the function of Hubble could be essentially more natural and convenient, than penetrability and evolution of shape of the wave packet.
To continue such a line, we perform some calculations of the duration of evolution of the universe defined by (\ref{eq.model.4.1})
(see Fig.~\ref{fig.11}, where we restrict ourselves by choosing $a_{\rm min} = a_{\rm start}$ in this paper).
\begin{figure}[htbp]
\centerline{\includegraphics[width=92mm]{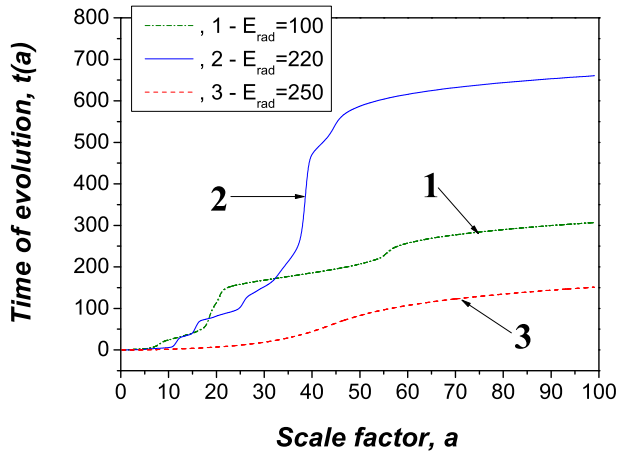}}
\vspace{-2mm}
\caption{\small 
Duration of evolution of the universe (its age) as function of the scale factor
(calculation parameters: the starting point $a_{\rm start} = 0.01$, 10000 intervals at $a_{\rm max} = 100$, $a_{\rm min}=a_{\rm start}$).
Model~\cite{Maydanyuk.2011.EPJP} at $k = 1$ (green dash-dotted line 1 for $E_{\rm rad}=100$, blue solid line 2 for $E_{\rm rad}=220$, red dashed line 3 for $E_{\rm rad}=250$).
For $a < 65$ larger duration at $E_{\rm rad} = 220$ in comparison with the duration at $E_{\rm rad} = 100$ is consistent with the previous results for the velocities in Fig.~\ref{fig.6}~(a,~b):
the velocity at $E_{\rm rad} = 100$ is changed more strongly, that causes more strong increase of the duration.
But for larger $a$ one can see similar increase of all durations, that corresponds to similar tendencies of the velocities to unite limit as $a \to 100$.
\label{fig.11}}
\end{figure}

At finishing, we support the presented above characteristics by classical calculations which the standard classical cosmology gives.
We have Friedmann equation:
\begin{equation}
\begin{array}{lll}
  H^{2} =
  \Bigl( \displaystyle\frac{\dot{a}}{a} \Bigr)^{2} =
  \displaystyle\frac{8\pi\, G}{3}\; \rho(a) -
  \displaystyle\frac{k}{a^{2}},
\end{array}
\label{eq.results.3.1}
\end{equation}
where energy density $\rho(a)$ is defined in (\ref{eq.model.1.2}).
Solving it, we find formula for velocity
\begin{equation}
\begin{array}{lll}
  v =
  \dot{a} =
  \displaystyle\frac{da}{dt} =
  \sqrt{\displaystyle\frac{8\pi\, G}{3}\: a^{2} \rho(a) - k}
\end{array}
\label{eq.results.3.2}
\end{equation}
and duration of existence of the Universe
\begin{equation}
\begin{array}{lll}
  t = t_{0} +
  \displaystyle\int\limits_{a(t_{0})}^{a(t)}
    \displaystyle\frac{da}{\sqrt{\displaystyle\frac{8\pi\, G}{3}\: a^{2} \rho(a) - k}}.
\end{array}
\label{eq.results.3.3}
\end{equation}
Results of such calculations at $E_{\rm rad}=100$ are presented in next Fig.~\ref{fig.11}.
\begin{figure}[htbp]
\centerline{\includegraphics[width=92mm]{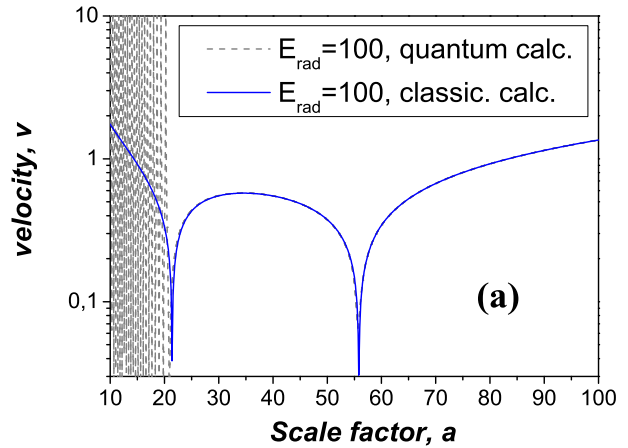}}
\vspace{-6mm}
\centerline{\includegraphics[width=92mm]{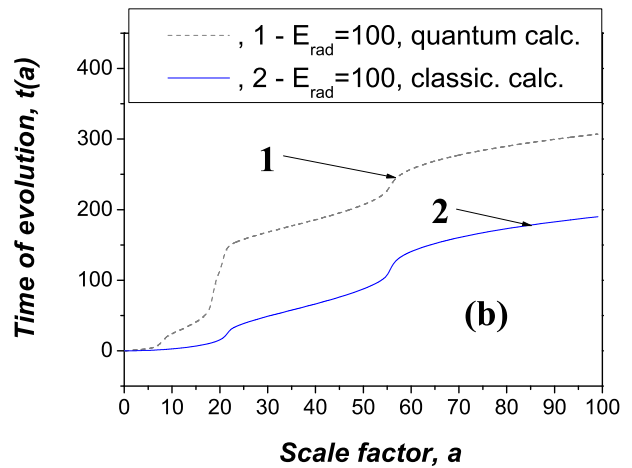}}
\vspace{0mm}
\caption{\small 
Classical calculations of velocity of evolution of the universe defined by (\ref{eq.results.3.2}) (a) and Hubble parameter defined by (\ref{eq.results.3.1}) (b) at $E_{\rm rad}=100$ as function of the scale factor $a_{\rm new}$ in comparison with quantum calculations of such characteristics given by (\ref{eq.model.3.9}) and presented in Fig.~10
(quantum calculation parameters: the starting point $a_{\rm start} = 0.01$, 10000 intervals at $a_{\rm max} = 100$, $a_{\rm min}=a_{\rm start}$).
After the first minimum, we achieve excellent agreement between classical and quantum calculations
of the velocity and Hubble parameter.
%
\label{fig.11}}
\end{figure}
From here one can see that we achieve excellent agreement between classical and quantum calculations after the first minimum at $a=21$.
Such an agreement confirms advance of the proposed our definitions (\ref{eq.model.3.6}), (\ref{eq.model.3.9}) and (\ref{eq.model.4.1}) for the velocity of expansion of the Universe, parameter of Hubble and duration of evolution of the Universe, which we introduce into quantum cosmology. At the same time, using such a correspondence between classical and quantum calculations, we can use quantum information for description of the initial stage before formation of the Universe with classical space-time (i.e. before Big Bang).
The classical curve for time has similar behavior as quantum, but there is shift, which can be connected by choosing of the starting moment $t_{0}$.


\section{Conclusions
\label{sec.conclusions}}

The formation of the universe of closed type and its further expansion in the first evolution stage are studied in the framework of Friedmann-Robertson-Walker metrics in the quantum consideration.
In order to form quantum basis for description of dynamics of evolution of the universe, we introduce operators of the velocity of expansion and the function of Hubble, and define duration of the evolved universe after its formation.
We demonstrate that the proposed definitions are characterized by high stability of calculations and easy for use.
In particular, such a way working with barriers of arbitrary shape is essentially more stable and effective
than other existed quantum dynamical approach based on evolution of the wave packet
(for example, see \cite{AcacioDeBarros.2007.PRD,Monerat.2007.PRD,%
Pedram.2007.IJTP,Pedram.2007.PLB,Pedram.2007.CQG,%
Pedram.2008.PLB.v659,Pedram.2008.PLB.v660,Pedram.2008.GRG,Pedram.2008.PRD,Pedram.2008.JCAP,%
Pedram.2009.PLB,Pedram.2010.IJTP,Pedram.2010.PLB,%
Correa_Silva.2009.PRD,Vakili.2010.CQG,Majumder.2011.PLB}
for comparison of results).
The introduced characteristics are supported by calculations of wave function in the fully quantum (non-semiclassical) approach.
Resonant influence of the initial and boundary conditions on the barrier penetrability is observed.
%
%
We achieve high precision agreement between the classical and quantum calculations after the formation of Universe with classical spacetime (i.e. Big Bang).
Such an agreement confirms efficiency of the proposed definitions, and classical-quantum correspondence allows to obtain quantum information before Big Bang, to study dynamics of evolution of universe in the first stage and later times.



\end{document}